\documentstyle[prl,aps,epsf]{revtex}

\begin{document}

\draft

\twocolumn[\hsize\textwidth\columnwidth\hsize\csname @twocolumnfalse\endcsname %

\title{Charge Transport in the Dense Two-Dimensional Coulomb Gas}

\author{Dierk Bormann \cite{e-mail}}

\address{
Institut f\"ur Physik der Universit\"at Augsburg, Memminger Str.\ 6, D-86135 %
Augsburg, Germany
}
\date{Replaced 20 May 1997; final version to be published in Phys.\ Rev.\ Lett.}

\maketitle

\begin{abstract}
The dynamics of a globally neutral system of diffusing Coulomb charges in two %
dimensions, driven by an applied electric field, is studied in a wide %
temperature range around the Berezinski\u\i-Kosterlitz-Thouless transition.
I argue that the commonly accepted ``free particle drift'' mechanism of charge %
transport in this system is limited to relatively low particle densities.
For higher densities, I propose a modified picture involving collective %
``partner transfer'' between bound pairs.
The new picture provides a natural explanation for recent experimental and %
numerical findings which deviate from standard theory.
It also clarifies the origin of dynamical scaling in this context.
\end{abstract}

]

The Coulomb gas in two dimensions (2D CG) is the generic model for topological %
excitations (``vortices'') in 2D systems with a U(1) order parameter symmetry %
\cite{Minnhagen-RMP}.
Some important examples are the {\em XY} model of 2D planar magnets %
\cite{BKT}, superfluid or superconducting films %
\cite{Minnhagen-RMP,BKT,Nelson}, Josephson junction arrays \cite{JJA} and, %
with some modifications, 2D melting \cite{Minnhagen-RMP,Nelson}.
Vortices form a CG due to the characteristic, logarithmic form of their %
interaction potential.
Furthermore, via a suitable choice of boundary conditions (an imposed current %
in the case of superfluid systems) a topological ``electric field'' $E$ can be %
applied on the vortices. 
For conceptual clarity, {\em we shall use the CG language throughout this %
article} \cite{remark_SC}.

The globally neutral 2D CG in $E = 0$ undergoes the famous Berezinski\u%
\i-Kosterlitz-Thouless (BKT) transition \cite{BKT}.
For temperatures $T < T_{\rm\scriptscriptstyle BKT}$, all particles are %
effectively bound in neutral pairs and the system is a dielectric insulator. 
At $T_{\rm\scriptscriptstyle BKT}$ the pairs start to thermally dissociate, %
leading to metallic screening and conduction. 
In 2D superfluid systems, this unbinding of vortices causes the transition %
between superfluid and normal phases.

Besides its equilibrium phase transition, also the {\em dynamical} behavior of %
the 2D CG is of considerable interest, since, e.g., in any superfluid system, %
moving vortices are the main cause of {\em dissipation}.
A few years after the BKT papers, a number of publications %
\cite{McCauley,Huberman+,Myerson,AHNS} therefore extended the BKT pairing idea %
to nonequilibrium situations, studying in particular the current response to a %
static, homogeneous applied field $E$. 
Their common procedure was to depart from the BKT decomposition of the system %
into two components, ``bound pairs'' and ``free'' particles, and to add %
kinetic equations for the particle exchange between both; other interactions %
than the intra-pair one are cast into (both dielectric and metallic) screening %
of the latter.
Ambegaokar {\em et al.}\ (AHNS) \cite{AHNS} extended the approach to an %
oscillating field $E$ and concluded that at high driving frequencies the %
response is dominated by the internal dynamics of individual bound pairs (as %
well as, above $T_{\rm\scriptscriptstyle BKT}$, thermally generated free %
particles) whereas at low enough or zero frequency it is essentially due to %
free particle drift only.
From BKT theory it is known \cite{Minnhagen-RMP} that below $T_{\rm%
\scriptscriptstyle BKT}$ the charge correlation function decays algebraically %
as $r^{-\beta(T)}$ with distance $r$ where $\beta(T) := 1/(\tilde{\epsilon}(T) %
T)$, $T$ being the CG temperature and $\tilde{\epsilon}(T)$ its static %
dielectric constant.
The assumption of pure bound-pair reponse leads to algebraic decay in time as %
well with a dynamical exponent $z=2$ at all temperatures $T \leq T_{\rm%
\scriptscriptstyle BKT}$ \cite{Ambegaokar+Teitel}.
Field-induced pair dissociation allows for a nonvanishing static response %
below $T_{\rm\scriptscriptstyle BKT}$: AHNS found a nonlinear current-voltage %
({\em IV}) relation $j \propto E^a$ ($j$ is the CG current density here %
\cite{remark_SC}) with
\begin{equation}
  a(T) = \beta(T)/2 + 1
\ .\label{eq:i1}\end{equation}
Right at the BKT transition, $\beta(T_{\rm\scriptscriptstyle BKT}^-) = 4$ such %
that $a(T_{\rm\scriptscriptstyle BKT}^-)=3$, which together with $z=2$ %
satisfies the dynamical scaling \cite{SC-scaling} relation $a=z+1$.

Away from $T_{\rm\scriptscriptstyle BKT}$ however, doubt has recently been %
cast onto relation (\ref{eq:i1}) by dynamical simulations of the classical {%
\em XY} model \cite{MWJO} as well as of the 2D CG in the continuous plane %
\cite{Holmlund+} and on a lattice \cite{Weber+}, by a re-analysis of earlier %
experiments on layered superconductors \cite{Weber+}, and also by new %
experiments on 2D proximity coupled Josephson junctions arrays \cite{Newrock+}.
A common conclusion of all of these investigations is that $a$ rather seems to %
obey the relation
\begin{equation}
  a(T) = \beta(T) -1
\ ,\label{eq:i2}\end{equation}
proposed by Minnhagen {\em et al.}\ (MWJO) \cite{MWJO}.
In other work \cite{Simkin+} better agreement with form (\ref{eq:i1}) is %
found, so the issue is controversial.
Right at $T=T_{\rm\scriptscriptstyle BKT}^-$, both expressions give a value of %
3, but (\ref{eq:i2}) has a steeper increase with decreasing $T$.
Assuming that the critical scaling relation $a=z+1$ holds in an extended %
region below $T_{\rm\scriptscriptstyle BKT}$, MWJO link (\ref{eq:i2}) to a %
temperature dependent dynamical exponent
\begin{equation}
  z(T) = \beta(T) - 2
\ ,\label{eq:i3}\end{equation}
providing further scaling arguments in favour of (\ref{eq:i3}).

In this letter, I propose a concrete physical picture for the correlated %
particle motion in a regime of {\em intermediate} CG particle density $n$, %
which {\em in a natural way leads to both relations (\ref{eq:i2}), (%
\ref{eq:i3}) independently.}
My picture is an extension of the ``conventional'' kinetic theory of Refs.\ %
\cite{McCauley,Huberman+,Myerson,AHNS} and, accordingly, reproduces their %
results (\ref{eq:i1}) and $z \equiv 2$ for the ``true'', asymptotic exponents %
in the limit of small $n$.
Furthermore, the underlying collective processes require a sufficiently strong %
applied field $E$ (in the case of d.c.\ nonlinear response), or a %
sufficiently high frequency $\omega$ (in the case of a.c.\ linear response), %
respectively. 
Consequently, at fixed, finite $n$ we also recover (\ref{eq:i1}) for small %
$E$, and $z=2$ for small $\omega$. 
The detailed crossover conditions are derived below; their direct observation %
in experiments or simulations is however lacking so far.

Following Refs.\ \cite{McCauley,Huberman+,Myerson,AHNS}, we consider %
overdamped particle motion, described by a system of Langevin equations
\begin{equation}
  \frac{{\rm d} {\bbox{x}}_i}{{\rm d} t} = \mu_0 q_i \left[ \sum_{j(\ne i)} {%
  \bbox{\nabla}}_i V_0(|{\bbox{x}}_i - {\bbox{x}}_j|) q_j + {\bbox{\xi}}_i + {%
  \bbox{E}} \right]
\ \label{eq:i4}\end{equation}
where ${\bbox{x}}_i$ are the CG particle positions and $q_i \in \{\pm1\}$ %
their charges ($\sum_i q_i = 0$).
$\mu_0$, $V_0$ are the ``bare'' particle mobility and Coulomb potential; $V_0$ %
behaves as $V_0(r) \sim \ln(r/r_0)$ for $r \gg r_0$, and $V_0(r) \sim 0$ for %
$r \ll r_0$ to prevent pair collapse \cite{Minnhagen-RMP}.
The heat bath is represented by delta-correlated gaussian random forces ${%
\bbox{\xi}}_i$.

Let me briefly recall the conventional ``free particle drift'' picture %
\cite{McCauley,Huberman+,Myerson,AHNS} of {\em IV} response to a static %
applied field $E$. 
The dynamically screened pair problem is described by a two-particle Langevin %
equation of the form (\ref{eq:i4}) with $\mu_0$ replaced by an effective %
charge mobility $\mu_{\rm eff}$, and $V_0$ replaced by the {\em statically} %
screened pair potential $V_{\rm scr}$ which we approximate as
\begin{equation}
  V_{\rm scr}(r) \approx \left\{ 
    \begin{array}{ll}
      V_0(r)/\tilde{\epsilon} & , \ r < \ell_{\rm scr} \\
      V_0(\ell_{\rm scr})/\tilde{\epsilon} & , \ r > \ell_{\rm scr} \\      
    \end{array}
  \right.
\ .\label{eq:i7}\end{equation}
$\ell_{\rm scr}$ denotes the metallic screening length due to free charges and %
$\tilde{\epsilon}$ the macroscopic, static dielectric constant of bound pairs %
up to this scale $\ell_{\rm scr}$.
Both $\mu_{\rm eff}$ and $\tilde{\epsilon}$ are actually weakly scale %
dependent quantities, but this does not change the conclusions in any %
essential way and we ignore it here.
$\mu_{\rm eff}$ only sets the overall timescale and thus does not enter the %
exponents $a$ or $z$.

Let us assume $T < T_{\rm\scriptscriptstyle BKT}$ (i.e., $\beta > 4$) for the %
moment, such that for $E=0$ all particles are bound in pairs and there is no %
metallic screening.
Under the action of $E \ne 0$, pairs dissociate and free particles recombine %
with finite rates.
Standard transition state theory yields for the respective rates per area $%
\Gamma_{\rm diss} \propto y E^\beta$ and $\Gamma_{\rm rec} \propto n_{\rm %
f}^2$, where $y$ is the fugacity of pairs at a distance $r_0$ which controls %
the total particle density and $n_{\rm f}$ is the density of free particles.
In the stationary state, both rates are equal.
Assuming then that the current is caused by free particle drift in the field %
$E$, we arrive at 
\begin{equation}
  j_{\rm dr} \propto n_{\rm f} E \propto y^{1/2}E^{\beta/2+1}
\ ,\label{eq:a2}\end{equation}
i.e., $a_{\rm dr} = \beta/2+1$ [compare Eq.\ (\ref{eq:i1})]. 

In order to see under which conditions the above argument holds, let us look %
at length scales in the problem which are relevant to transport.
One is the separation $\ell_E = 1/\tilde{\epsilon} E$ at which a pair becomes %
unstable in the applied field $E$; pairs smaller than $\ell_E$ are %
(temporarily) bound.
Another is the typical distance $\ell_{\rm dr}$ which a free particle drifts %
in the direction of $E$ until it recombines, 
\begin{equation}
  \ell_{\rm dr} := j_{\rm dr}/2\Gamma_{\rm rec} = j_{\rm dr}/2\Gamma_{\rm %
  diss} \propto y^{-1/2}E^{-\beta/2+1}
\ .\label{eq:a3}\end{equation}
Since a neutral pair of size $<\ell_E$ is regarded as bound, a free particle %
density $n_{\rm f} > \ell_E^{-2}$ or, equivalently, a ``drift length'' $\ell_{%
\rm dr} < \ell_E$ clearly would be physically meaningless. 
Relations (\ref{eq:a2}) and (\ref{eq:a3}) imply $n_{\rm f}\ell_E^2 \propto %
\ell_E/\ell_{\rm dr} \propto y^{1/2}E^{\beta/2-2}$, so for sufficiently small %
$y$ or $E$ there is no contradiction.
With increasing $y$ and $E$, however, $n_{\rm f}\ell_E^2$ and $\ell_E/\ell_{%
\rm dr}$ must exceed $1$ at some point, implying a breakdown of the ``free %
particle drift'' picture and a crossover to a new ``partner transfer'' (PT) %
regime, where the constituents of a dissociated pair essentially have no %
opportunity anymore to drift freely, but immediately recombine with others.
The PT regime may be visualized as a dense liquid of close-to-critical pairs %
of size $\approx \ell_E$ (by definition these cannot ``overlap'' each other), %
``transferring partners'' among each other by thermally activated hopping %
\cite{remark_Petschek+Zippelius}.
The whole is of course immersed in a screening medium of smaller pairs, as %
discussed above.

In order to estimate the current in the PT regime, let us focus on the typical %
distance $\ell$ which a given particle separating from its partner is dragged %
along by $E$ until it binds to the next.
This length shrinks from $\ell \sim \ell_{\rm dr} \gg \ell_E$ in the ``drift'' %
regime to its lower bound $\ell_{\rm\scriptscriptstyle PT} \sim x_{\rm%
\scriptscriptstyle PT} \ell_E$ (with a dimensionless constant $x_{\rm%
\scriptscriptstyle PT} \agt 1$) in the PT regime. 
Assuming then that pair dissociation in the PT regime is still approximately %
described by transition state theory (implying $\Gamma_{\rm diss} \propto y E^%
\beta$), we find
\begin{equation}
  j_{\rm\scriptscriptstyle PT} = 2\Gamma_{\rm diss} \ell_{\rm%
\scriptscriptstyle PT} \propto y E^{\beta-1}
\ ,\label{eq:a5}\end{equation}
i.e., $a_{\rm\scriptscriptstyle PT} = \beta-1$ precisely as obtained in Refs.\ %
\cite{MWJO,Holmlund+,Weber+} [see Eq.\ (\ref{eq:i2})]!
Note that an extension of the Kosterlitz renormalization procedure by %
including $E$ as a scaling variable \cite{Myerson,Sujani+} does {\em not} have %
a comparable effect, which indicates that the PT picture captures additional %
aspects of the strongly correlated dynamics at higher density.

MWJO \cite{MWJO} have attempted to reconcile the new relation $a_{\rm%
\scriptscriptstyle PT} = \beta-1$ with the ``free particle drift'' picture, {%
\em postulating} a modified dependence $\Gamma_{\rm rec} \propto n_{\rm %
f}^{1+2/(\beta-2)}$ of the recombination rate on the free particle density $n_{%
\rm f}$. 
If however, as we have argued here, the concept of free particles itself %
becomes questionable in the new regime, this type of explanation clearly %
cannot be adequate. 

In the high temperature phase ($\beta<4$) thermal pair dissociation leads to %
metallic screening, which provides a further characteristic length %
\cite{Minnhagen-RMP}
\begin{equation}
  \ell_{\rm scr} \propto n_{\rm f}^{-1/2} \propto y^{-1/(4-\beta)}
\ .\label{eq:a6}\end{equation}
As long as $\ell_{\rm scr} > \ell_E$, field induced pair dissociation %
dominates over thermal one and our arguments remain essentially valid, only %
the order of the ``drift'' and PT regimes is reversed: the PT regime now is at %
{\em small} $E$ (but still intermediate $y$).
At still smaller $E$ such that $\ell_{\rm scr} < \ell_E$ one eventually %
obtains an ohmic regime.

The above results for $j(y,E)$ can be expressed by a scaling form of the %
nonlinear CG conductivity $\sigma_{\rm\scriptscriptstyle CG}$ \cite{remark_SC},
\begin{equation}
  \sigma_{\rm\scriptscriptstyle CG}(y,E) := j(y,E)/E = \ell_E^{-2} \widetilde{%
  \sigma}(\ell_E/\ell_{\rm dr},\ell_E/\ell_{\rm scr})
\ ,\label{eq:a9}\end{equation}
where the scaling function $\widetilde{\sigma}$ depends on $\beta$ (but not %
$y$) and has the asymptotic forms $\widetilde{\sigma}(u,0) \propto u$ for $u %
\ll 1$ (``drift'') and $\propto u^2$ for $u \gg 1$ (PT), respectively, and $%
\widetilde{\sigma}(u,v) \propto v^2$ for any $u$ and $v \gg 1$ (ohmic regime).

\begin{figure}
\vspace*{-5mm}
\epsfxsize=70mm
\centerline{\epsffile{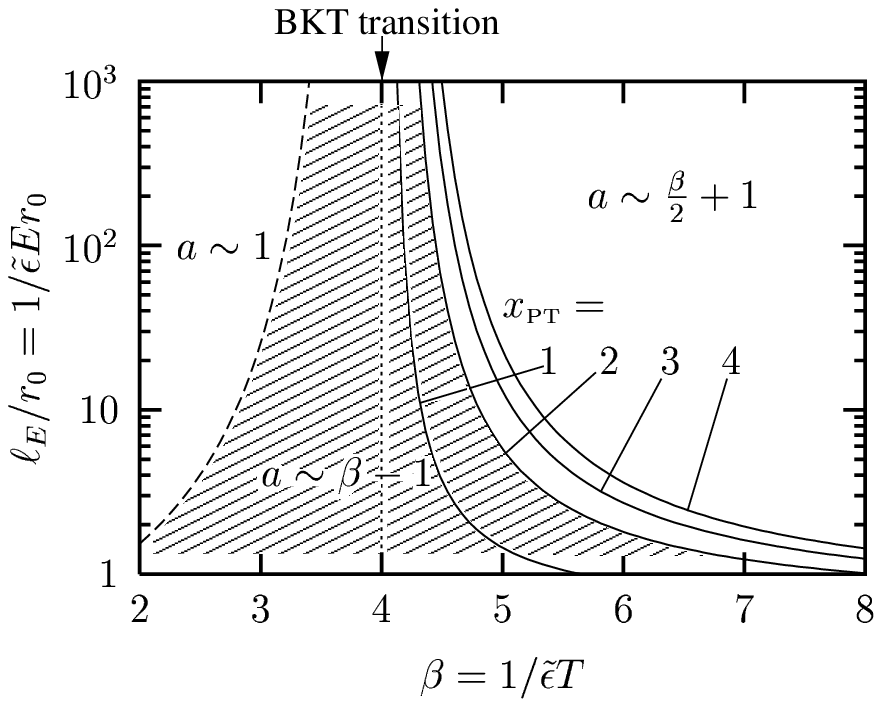}}
\caption{
Different {\em IV} regimes in the $\beta$-$E^{-1}$ plane according to Eqs.\ (%
\ref{eq:b1}), (\ref{eq:b2}) and using the ``Villain'' relation $\ln y = -\beta %
\pi/2$, for some reasonable values of $x_{\rm\scriptscriptstyle PT}$.
The PT regime, characterized by the condition $\ell_{\rm dr}/x_{\rm%
\scriptscriptstyle PT} < \ell_E < \ell_{\rm scr}$, expands with increasing $x_{%
\rm\scriptscriptstyle PT}$ (shading is for $x_{\rm\scriptscriptstyle PT} = 2$).
}
\label{fig:E_cro}
\end{figure}

Estimating quantitatively the prefactors in Eqs.\ (\ref{eq:a3}) and (%
\ref{eq:a6}), we find that the new PT regime extends to relatively small $y$ %
and $E$, so its predominance in experiments and computer simulations is not %
surprising.
Using transition state theory to consistently estimate both pair dissociation %
\cite{AHNS} and recombination \cite{McCauley} rates we obtain
\begin{equation}
  \frac{\ell_{\rm dr}}{r_0} \approx \frac{\,{\rm e}^{-\beta/2}}{\sqrt{8\pi y}} %
  \left( \frac{\ell_E}{r_0} \right)^{\beta/2-1}
\ .\label{eq:b1}\end{equation}
As to the screening length $\ell_{\rm scr}$, a simple Debye-H\"uckel estimate %
\cite{Minnhagen-RMP} turns out to be quantitatively dissatisfactory since it %
implies $\ell_{\rm scr} < n_{\rm f}^{-1/2}$ at $\beta \approx 4$.
Minimizing the grand canonical potential of a mixed system of bound pairs and %
free particles with respect to $\ell_{\rm scr}$ yields instead %
\cite{remark_critical}:
\begin{equation}
  \frac{\ell_{\rm scr}}{r_0} \approx \left( \frac{2\pi}{\beta} \sqrt y %
  \right)^{-2/(4-\beta)}
\ .\label{eq:b2}\end{equation}
Equations (\ref{eq:b1}), (\ref{eq:b2}) imply an extension of the PT regime to %
arbitrarily large $\ell_E$ (i.e., small $E$) in the vicinity of the BKT %
transition, since $\ell_{\rm scr}$ diverges as $\beta \to 4^-$ and, for $8\pi %
y > \,{\rm e}^{-4}$ and any finite $\ell_E$, $\ell_{\rm dr}/\ell_E$ shrinks to %
below $1$ as $\beta \to 4^+$.
For typical 2D vortex systems \cite{MWJO,Weber+,Newrock+}, $y$ and $\beta$ %
obey to a good approximation the ``Villain'' relation $\ln y \approx -\beta %
\pi/2$ \cite{Minnhagen-RMP}. 
Figure \ref{fig:E_cro} shows the resulting location of the different regimes %
in the $\beta$-$E^{-1}$ plane for some reasonable values of the (so far %
unknown) constant $x_{\rm\scriptscriptstyle PT}$: in particular, in the %
vicinity of the BKT transition the asymptotic exponent $a_{\rm dr} = \beta/2 + %
1$ appears only at extremely small driving fields $E$ which probably are not %
reached in these investigations.

\begin{figure}
\vspace*{-5mm}
\epsfxsize=85mm
\centerline{\epsffile{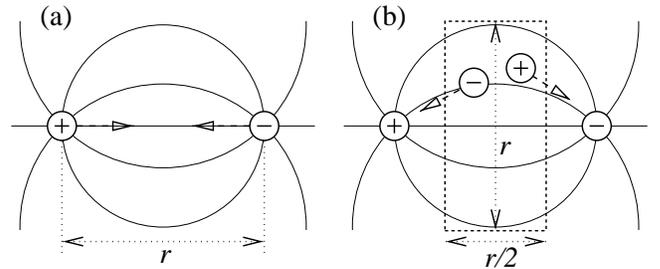}}
\caption{
Pair equilibration by (a) ``drift'' and (b) ``partner transfer'' (PT).
In (b), the relevant region of small-pair dissociation, leading to PT %
relaxation of the big one, is framed by a dashed line.
Electric field lines of the big pair are shown. 
}
\label{fig:relax}
\end{figure}

We now turn to dynamical correlations at $E=0$.
Following Refs.\ \cite{McCauley,Huberman+,Myerson,AHNS}, we assumed here that %
the motion of a given bound pair or free particle is essentially uncorrelated %
with the rest of the system, feeling it only through its screening properties.
In particular, at $T<T_{\rm\scriptscriptstyle BKT}$, charge density %
correlations are governed by the internal dynamics of pairs %
\cite{Ambegaokar+Teitel}.
The effective dynamical exponent $z$ for fluctuations with wavenumbers around %
$k$ can then be read off from the typical relaxation time $\tau(r) \propto %
r^z$ of pairs of size $r \propto k^{-1}$; it thus will depend on whether the %
relaxation process is dominated by ``drift'' or PT.
The two mechanisms are illustrated in Fig.\ \ref{fig:relax}: (a) a member of a %
pair can simply drift towards the other under the action of its field, or (b) %
each of them can recombine with a member of another, smaller pair sitting in %
between them.
In case (a) each particle has to move a distance $\propto r$ pulled by the %
field $E_{\rm P} \propto 1/r$ of its partner, which implies
\begin{equation}
  \tau_{\rm dr}(r) \propto r/E_{\rm P} \propto r^2
\ ,\label{eq:a7}\end{equation}
i.e., $z_{\rm dr}=2$.
This relaxation mechanism has been analyzed in detail by Ambegaokar and Teitel %
\cite{Ambegaokar+Teitel} and is certainly the dominant one in the limit of %
large $r$.
In case (b) there is an area $\propto r^2$ within the big pair in which its %
dipole field has a strength $E_{\rm D} \propto 1/r$ and thus induces the %
dissociation of smaller pairs with a rate per area $\Gamma_{\rm diss} \propto %
yE_{\rm D}^\beta$, so the typical waiting time until such an event occurs %
behaves as \cite{neutralizing_PT_current}
\begin{equation}
  \tau_{\rm\scriptscriptstyle PT}(r) \propto 1/r^2\Gamma_{\rm diss} \propto %
  y^{-1}r^{\beta-2}
\ ,\label{eq:a8}\end{equation}
i.e., $z_{\rm\scriptscriptstyle PT}=\beta-2$ in agreement with (\ref{eq:i3}).
The {\em four-body} correlations implicit in this argument again underline the %
collective character of the PT mechanism.
Comparing both relaxation times, $\tau_{\rm dr}(r)/\tau_{\rm\scriptscriptstyle %
PT}(r) \propto yr^{4-\beta}$, shows that below the BKT transition ($\beta>4$) %
the ``drift'' mechanism is dominant at small $y$ or $k$ ($\propto r^{-1}$), %
whereas with increasing $y$ and $k$, PT takes over at some point. 

(\ref{eq:a7}), (\ref{eq:a8}) imply dynamical scaling forms of the charge %
density correlation function which in turn, via the fluc\-tu%
\-ation-dissipation theorem, suggest a dynamical extension of (\ref{eq:a9}) to %
$\sigma_{\rm\scriptscriptstyle CG}(y,E,\omega)$, including a further scaling %
variable $\ell_{\rm dr}/\ell_\omega$ ($\ell_\omega := \sqrt{\mu_{\rm eff} T/%
\omega}$).
For sufficiently large $y$ and $\omega$, the PT mechanism dominates, leading %
to a simplified form $\sigma_{\rm\scriptscriptstyle CG}(y,E,\omega) \sim \ell_%
\omega^{-2} \widetilde{\sigma}'(\ell_E^{-z_{\rm\scriptscriptstyle PT}}/\ell_%
\omega^2)$ (the form of the scaling variable reflects the fact that $\tau_{\rm%
\scriptscriptstyle PT}(\ell_E) \propto \ell_{\rm dr}^{-2} \propto \ell_E^{-z_{%
\rm\scriptscriptstyle PT}}$ in the PT regime).
The validity of the ``critical'' scaling relation $a_{\rm\scriptscriptstyle %
PT}=z_{\rm\scriptscriptstyle PT}+1$ in a whole $T$ {\em region} below $T_{\rm%
\scriptscriptstyle BKT}$ (which was an {\em assumption} in the MWJO scaling %
argument!) now follows very naturally by the usual reasoning \cite{SC-scaling}.
It expresses the fact that in the PT regime both dc charge transport and the %
decay of charge fluctuations are governed by the {\em same} mechanism, %
characterized by a {\em single} relevant length scale $\ell_E \propto 1/E$.

In summary, I have described a new dynamical regime of the 2D CG in which %
charge transport predominantly occurs through direct particle transfer among %
bound pairs, instead of free-particle drift.
It offers a natural interpretation of recent experimental and numerical %
findings on 2D vortex dynamics which deviate from ``standard'' AHNS theory %
\cite{AHNS}.
In contrast to the more extreme view held by MWJO \cite{MWJO}, I {\em do}, %
however, recover AHNS behavior asymptotically for low driving fields and %
frequencies; the new picture thus resolves the apparent contradiction between %
the MWJO and AHNS scenarios.

Further tests of this picture in experiments and numerical simulations are %
clearly needed; in particular, a direct observation of the predicted %
crossovers would be desirable.  
Another testable signature is the changed dependence of dc response (%
\ref{eq:a5}) and relaxation times (\ref{eq:a8}) on the pair fugacity $y$.  
I am not aware of any controlled way to tune $y$ independently of $\beta$ in %
real superconducting systems.  
In experiments on superfluid $^4$He films, the vortex fugacity can be {\em %
increased} above the ``Villain'' value by adding $^3$He impurities, which %
stick to the vortex cores and lower their energy \cite{4He/3He}.
However, as we have seen, in order to observe the PT/``drift'' crossover a {%
\em reduction} of the fugacity would be needed instead.

In any event, in real systems a reliable determination of the effective CG %
temperature $T$ and thus of $\beta$ usually is an extremely difficult task.  
In contrast, CG simulations are natural test candidates since here $T$, $y$ %
are separately controllable model parameters (within certain limits this is %
also true for generalized {\em XY} type models \cite{Jonsson+Minnhagen}).  
Preliminary Monte Carlo data for a lattice CG \cite{Weber_personal} indeed %
support the proportionality (\ref{eq:a5}) between $j$ and $y$ in the %
accessible parameter range.

The crossover at low $E$ \cite{remark_SC} in the {\em IV} characteristics %
should in principle be observable both in experiments and simulations.  
Figure \ref{fig:E_cro} suggests to look for it at temperatures sufficiently %
far below $T_{\rm\scriptscriptstyle BKT}$ ($\beta \agt 5$, say), since the %
crossover value of $\ell_E$ diverges at $T_{\rm\scriptscriptstyle BKT}$.
The same is true in principle for the low $\omega$ crossover in the dynamical %
correlations (i.e., in the ac linear response) although this one may be covered %
by other collective effects \cite{Jonsson+Minnhagen,Capezzali+}.

I would like to thank P. Minnhagen, K. Holmlund, M. Nylen, O. Westman, H. %
Weber, H. Beck and S. R. Shenoy for clarifying conversations.
This work was supported by grants from `Svenska Forskningsr{\aa}det' and %
`Deutsche Forschungsgemeinschaft'.



\begin{references}
\bibitem{remark_SC} 
For superconductors, the conventional quantities are as follows: (dc response) %
$I \propto E$, $V \propto j$; and (ac response) $\sigma(\omega) \propto [2\pi %
\sigma_{\rm\scriptscriptstyle CG}(\omega) - {\rm i}\omega]^{-1}$.
\bibitem{remark_critical}
Note that this form encloses the correct critical behavior $\ell_{\rm scr} %
\propto \,{\rm e}^{{\rm const}/\sqrt{T-T_{\rm\scriptscriptstyle BKT}}}$ %
because of the square root dependence $\tilde{\epsilon}(T) - \tilde{%
\epsilon}(T_{\rm\scriptscriptstyle BKT}) \propto \sqrt{T-T_{\rm%
\scriptscriptstyle BKT}}$ (see, e.g., \cite{Minnhagen-RMP}).
\end{references}
\end{document}